# Performance Enhancement of Ad Hoc Networks with Janitor Based Routing

Isnain Siddique, Rushdi Shams, and M.M.A. Hashem
*Department of Computer Science and Engineering*
*Khulna University of Engineering and Technology (KUET)*
*Khulna- 920300, Bangladesh*
Isnain_cse2k1@yahoo.com, rushdecoder@yahoo.com, mma_hashem@hotmail.com

## Abstract

*We propose and analyze a new on the fly strategy that discovers, repairs and maintains routes in hierarchical and distributed fashion called Janitor Based Routing (JBR). The main motivation behind our JBR protocol is to decrease flooding and routing overhead and increase efficiencies in packet movement. An analytical model for the proposed JBR is presented and detailed simulation is used to observe the performance of JBR. This route discovery and maintenance protocol clearly achieved improvement in terms of reduction of flooding, routing overhead, and, hence, provides enhanced reliability.*

## 1. Introduction

An ad hoc network is a class of wireless systems that consists of independent mobile nodes communicating with each other over wireless links, without any static infrastructure such as base stations. Two nodes that are out of each other's radio range communicate through an intermediate node, which also serves as a router. Since the nodes move randomly, the topology of the network changes with time. Dynamically changing topology and lack of centralized control make the design of an adaptive distributed routing protocol challenging. Many protocols have been proposed for mobile ad hoc networks, with the aim of making the route discovery and route maintenance more efficient [1].

Today, ad hoc networks are becoming popular because of their 3 "Anys"- Any person, Any where and Any time. But the problem of designing an efficient routing protocol for wireless systems has been studied by many researchers. Defense Advanced Research Projects Agency (DARPA) supported Packet Radio Network (PRNET) [2] and Survivable Radio Network (SURAN) [3] projects provide automatic route setup and maintenance in a packet radio network with moderate mobility. Recent interests in such networks have resulted in the formation of a working group within the Internet Engineering Task Force (IETF) called Mobile Ad hoc Networking (MANET). This group supports development of new routing protocols for ad hoc networks apart from tailoring wire line Internet protocols to ad hoc networks.

The routing protocols for MANET can be classified as proactive and reactive [4], depending on how they maintain routing information and how they respond to topology changes. These protocols, like Clusterhead Gateway Switch Routing CGSR [5], Zone Routing Protocol ZRP [6] or Localized Route Repair LRR [7] if they work efficiently, then they represent drawbacks in contrast. They surmise that they will avoid flooding but in essence, the network overhead increases formidably. The protocols can be categorized under many characteristics such as the structure, number of routes, possibility of source routing, route recover mechanisms, beckoning requirements, stored information, update period, update information, update destination and packet propagation method, etc. In this JBR protocol, we have proved it to be stoic in avoiding flooding. Proposed JBR protocol also reduces packet and byte overhead which is elaborated in the analytical model and simulation performance presented in this paper. On the fly route discovery algorithms suffer from the complexity, JBR is immune from them. JBR does not consume limited network bandwidth also. Nevertheless, measurement of efficiency in performance criteria is well beyond from the ad hoc network's point of view, which all other protocols promise, eventually stays miles away from JBR.

## 2. Related works

Many reactive and proactive protocols including a mixture of both approaches are concerned with efficient route discovery and route maintenance.

CGSR [5] was developed at University of California, Los Angeles UCLA in 1996, and just like

Destination-Sequenced Distance-Vector DSDV, it has been simulated but never implemented. Some of the key features of the protocol are that it uses a cluster head, code separation between the clusters and cluster-based channel access and routing. One limitation of this protocol is that it is based on DSDV as the underlying route update method, which can cause problems. The other limitation is that it uses periodic route and cluster membership updates. The protocol is based on the concept of clusters and cluster-heads. Routing is done via the cluster-heads and gateways. Data from a host is routed in such a way that it is sent to the nearest cluster-head, which then forwards that to a gateway node, which then sends it to the next cluster-head. The cluster member table is broadcasted periodically in order to have up-to-date information about the clusters. ZRP [6] was developed at Cornell University in 1998. Even though the protocol is not really specified, just the algorithms, simulations have been made in 1998. There is not any implementation known yet. The algorithm combines the proactive and reactive approaches and is built upon the concept of zones. So in this way ZRP uses table-driven routing for nodes within a routing zone (this is also called IntrAzone Routing Protocol - IARP) and on-demand query for nodes outside a routing zone (IntErzone Routing Protocol - IERP). Every node defines a zone radius, but the problem is that it is hard to decide upon an appropriate zone radius which is good for all applications. A newly developed approach LRR [7] uses a request zone, which is a zone in which route request packet propagates. In each intermediate node, NN (Next to Next) node information is embedded. If any node (upstream) finds route broken, it finds another node in contact with itself and NN node on the route.

## 3. Proposed Janitor Based Routing (JBR) protocol

The techniques in route discovery and route maintenance depending upon the zone, invokes a problem about the selection of the radius. Also, in addition, the code translations between zones create skewing network overheads. The similar problem exists in the case of pruning, albeit some efforts conceived the network overheads. Choosing an adept underlying protocol is another mammoth decision. Clusters often create localized overhead inaccessibility problem. It also has a tedious task of inter zonal route information translations.

The JBR uses nodes which are called janitors, with the maximum connectivity over the whole network. If one node gets alive, it broadcasts a message named "hello". This "hello" message is not periodic, rather it is event driven. Any node, getting this message from another node should give back a reply to that node only (not a broadcast). Detecting of janitor is completely individual responsibility and it is done in every node getting the information of the network. If one node has calculated that it has now become a janitor which was not in the past then it will send "new janitor" and all its covered nodes will accept it and the "hello" propagation immediately ends as it is no longer an ordinary node, it has become a janitor. The purpose of a network is to send or receive data so that the janitor available messages can be piggy backed with any data or acknowledge packet that has gone to janitor from that node or vice versa. Any node that has not sent any data to janitor and has not got any data from janitor over a predefined time, then to avoid complicacy, every node must inform its janitor that it is in its zone by a periodic message named "janitor alive request". This message does not continue in the "hello" session, in fact, when a "hello" session starts it stops and starts when the "hello" session stops with the information given by the "hello" session. If no reply of the "janitor alive request" is received by any node then it starts a "hello" session again. If the janitor does not receive piggy backed information within a predefined amount of time and no active data delivery is in that session, then it starts a new "hello" session.

In this approach, there are couple of cases that are needed to be considered.

1. If the destination (D) is directly connected with the source (S) (Fig. 1) then the data is simply sent to destination.

2. If the destination is not directly connected then the following cases may appear

(a) Source node sends data to its janitor (J). Janitor will check its connection. If it finds a direct connection with the destination the janitor will forward the data on behalf of the source (Fig. 2).

(b) If janitor does not get any direct connection with the destination, then it must look for the route in its cache and if such a route exists, it delivers data towards that way (Fig. 3).

(c) If a route gets broken dynamically, the node currently has the packet, backs the packet to the source with a "route error" message. So the "route recovery" packet is invoked and by this time that particular network is running the "hello" session (Fig. 4).

(d) If such a route does not evolve in the route cache of the janitor, it propagates queries to all the directly connected janitors ($J_1$, $J_2$, $J_3$) of it about the destination node. This process continues upwards. If any of them has it, then they may follow the reverse path to reply the query made by the janitor. If multiple paths are found then the janitor takes the path from which the reply can first. It keeps the path in its cache (Fig. 5).

(e) If a packet travels up to a predefined hop count and does not find its destination, then a route to the destination is estimated as not found. Then a “route unreachable” message is propagated by janitors (Fig. 6).

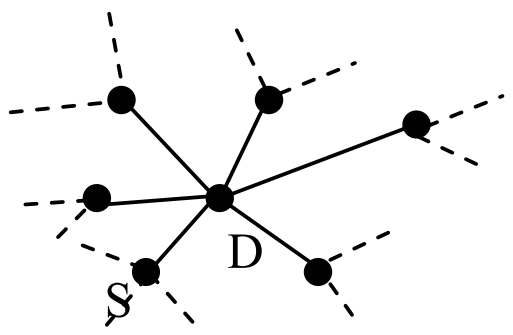

**Figure 1: Source is directly connected to destination**

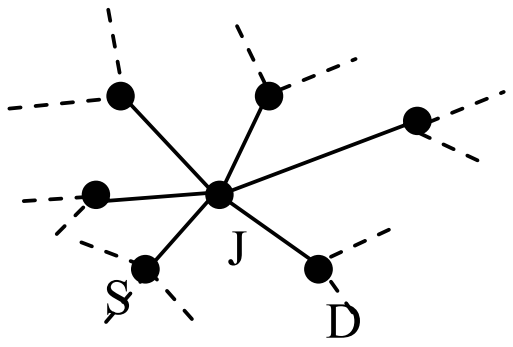

**Figure 2: Source node is not directly connected to destination node but janitor does**

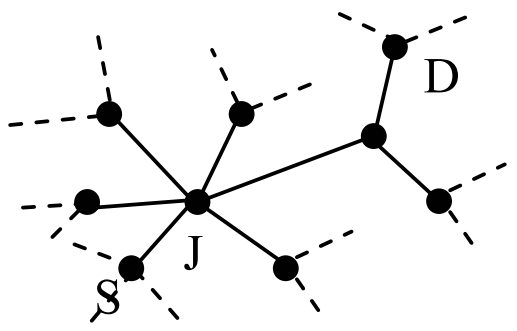

**Figure 3: Janitor has the destination its cache**

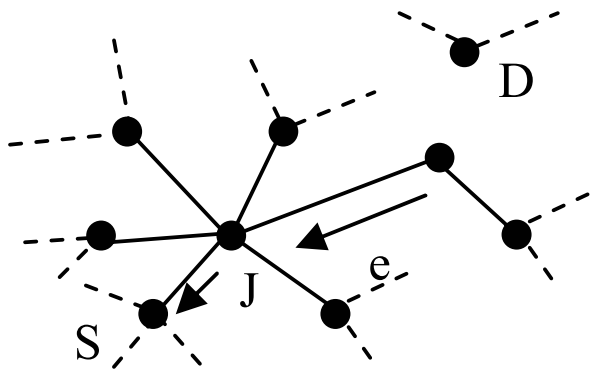

**Figure 4: Propagation of “route error” message**

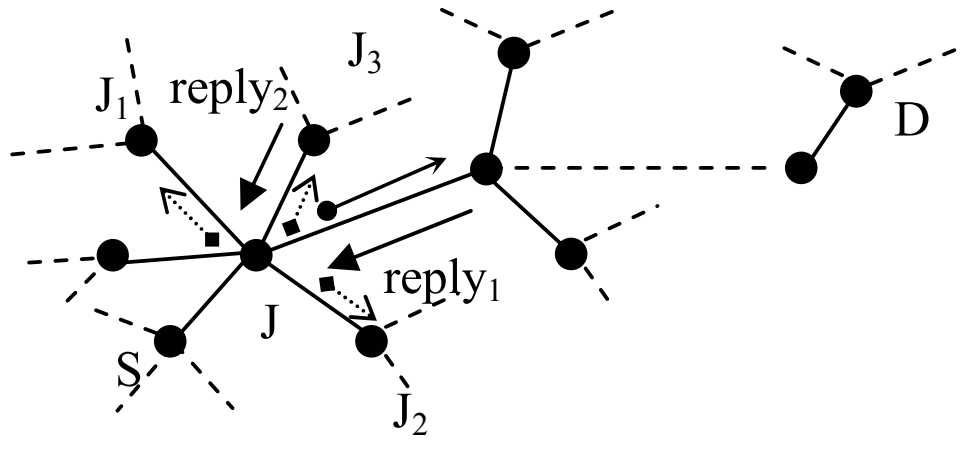

**Figure 5: Route query made by janitor**

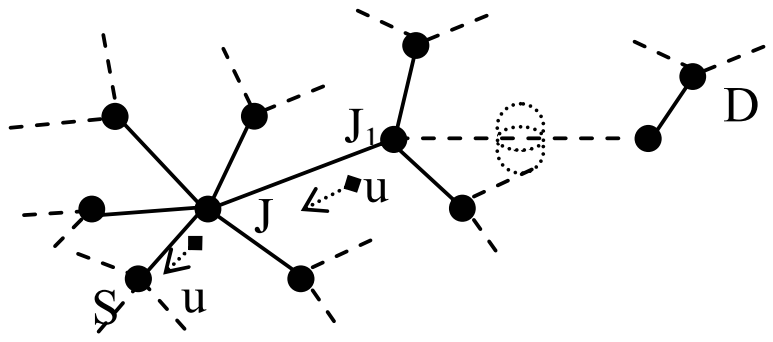

**Figure 6: Janitor forwards “route unreachable” message**

## 4. Analytical model

To develop a detailed model of the system and analyze the performance of JBR protocol, we assume certain parameters and arrive at an expression for traffic in ad hoc networks.

### 4.1. System model

We assume that there are $n$ nodes in the system, all the nodes have the same distribution of moving speed and direction and the same transmission range $r$. We assume that:

1. The average route length between the source and destination is $E_L$
2. The duration of the packet arrival is an exponentially distributed with mean $1/\lambda$ .
3. The time between location changes for each node is exponentially distributed with mean $1/\mu$ .

Then, probability of a route is broken [7],

$$P_B = \frac{\mu}{(\mu + \lambda)}$$

And the probability that a route is not broken is $(1 - P_B)$ .

*Lemma 1.* If there are exactly two nodes and they are within their range then after the hello session they are janitor of each other.

*Proof.* Let the two nodes be $m$ and $n$ and $\lambda$ is their number of active connection. So, $\lambda(m) = \lambda(n) = 0$. After the "hello" session, $\lambda(m) = \lambda(n) = 1$. So, they are janitor of each other.

*Lemma 2.* If janitor $J$ does not get any direct connection with the destination $d$, then it must look for the route in its cache and if such a route exists, it delivers data towards that way- this is true if $\lambda(J) > \lambda(s)$ and not necessarily $\lambda(J) > \lambda(d)$

*Proof.* The statement is true if and only if $J$ is the janitor of $S$. But eventually even the node $d$ can be itself a janitor. So the lemma holds.

*Lemma 3.* Not necessarily but every node has a possibility of being a janitor, depending on the network state.

*Proof.* Let, a node $m$ is connected with node $l$ and node $n$. if $\lambda(n) > \lambda(m)$ and $\lambda(l) > \lambda(n)$ then l is the janitor of $m$.

*Corollary 1.* If $\lambda(n) > \lambda(m)$ and $\lambda(l) > \lambda(m)$ and $m$ has no other connections, i.e. $\lambda(m) = 2$, then $m$ is not a janitor if $\hat{\lambda}_m(n) > 2$ and $\hat{\lambda}_m(l) > 2$. Here, $\hat{\lambda}_m(X)$ is the number of active connection of $X$ $_x$ except $m$.

## 4.2. Transmission cost analysis

*Theorem 1.* The probability of the packet for routing successfully is,

$$P_{RS} = \sum [P_L + P_{JS} - (P_L \vee P_{JS})]$$

*Proof.* If $P_L$ is the probability that a packet is successfully transmitted over a link and $P_{JS}$ is the probability that a packet has successfully finds its route by janitor, then the probability of the packet for routing successfully is,

$$\begin{aligned} P_{RS} &= \sum [P_{JS} \wedge P_L \ ] \\ &= \sum [P_L + P_{JS} - (P_L \vee P_{JS})] \end{aligned}$$

*Theorem 2.* $P_s$ is the probability that a packet is successfully routed to its final destination, then the average number of routing failures for a single packet is,

$$S = \frac{P_s}{E_L + (E_L - k) + (E_L - \hat{k})}$$

*Proof.* Let, $S$ be a random variable that describes the average number of routing attempts needed to successfully deliver the packet to its final destination. $k$ is the number of hops encountered before reaching to the janitor and $\hat{k}$ is the number of hops encountered after the janitor takes control. Then the random variable $S$ has the expectation,

$$E[S] = \sum_{k,\hat{k} \in E_L} SP(S)$$

*Corollary 2.* Let, $F$ be a random variable that describes the average number of routing attempts failed deliver the packet to its final destination.

$$F = \frac{(1 - P_s)}{E_L + (E_L - k) + (E_L - \hat{k})}$$

Then, the random variable $F$ has the expectation,

$$E[F] = \sum_{k,\hat{k} \in E_L} FP(F)$$

*Theorem 3.* If $C_R$ is the average cost of routing a packet to its final destination and if $C_{LS}$ is the cost of a successful link transmission, $C_{LF}$ is the cost of link failure, then the average cost of routing a packet to its final destination is

$$C_R = \begin{bmatrix} E_L + (E_L - k) \\ + (E_L - \hat{k}) + \dfrac{P_L(1 - P_S)}{P_S(1 - P_L)} \end{bmatrix} C_{RF} C_{RS}$$

*Proof.* We can observe that $C_R = zC_{RF}C_{RS}$, where $C_{RF}$ is the cost of route failure and $C_{RS}$ is the cost of error free routing. If the cost of route unsuccessful is $C_{RU}$ and the cost of drop of a query is $C_{QD}$, then the cost of routing failure is determined by the cost of partially routing the packet up to the broken link and the cost of information the sender about the error.

$$C_{RF} = qC_{LS} + qC_{LF} + qC_{QD} + qC_{RU}$$

The cost of error free routing depends on the route length $E_L$ and $C_{LS}$ of link transmission, thus we have,

$$\begin{aligned} C_{RS} &= E_L C_{LS} + (E_L - k)C_{LS} + (E_L - \hat{k})C_{LS} \\ &= E_L C_{LS} + E_L C_{LS} - kC_{LS} + E_L C_{LS} - \hat{k}C_{LS} \\ &= 3E_L C_{LS} - C_{LS}(k + \hat{k}) \end{aligned}$$

Then,

$$\begin{aligned} C_R &= z[q(C_{LS} + C_{LF} + C_{QD} + C_{RU})] \\ &[3E_L C_{LS} - C_{LS}(k + \hat{k})] \end{aligned}$$

## 4.3. Packet routing probabilities

*Theorem 4.* The probability $P_w$ that at least one of the $E_N$ janitors is able to route from source to destination is,

$$P_w = (1-(1-P_B)^3)^{E_N}$$

*Proof.* Let the packet is sent from node $A$ to $C$, where $C$ is not directly connected with $A$. So, an on demand diagnosis approach invokes a janitor $J$. The probability of $J$ to find the desired route is $\tau = (1-P_B)^3$. If $E_N$ be the number of janitors in the network, then $H$ be the number of nodes that want a route by janitor $J$ is given by,

$$P(H=K) = ({}^{E_N}C_K)\tau^K (1-\tau)^{E_N-K}$$

Thus, the probability that at least one of the $E_N$ janitors is able to route from source to destination is given by above.

### 4.4. JBR probabilities

In JBR protocol, a composite link transmission succeeds if and only if a route can be prescribed by the janitors. This means, the packet traverses $k$ hops initially to request to janitor. Therefore, the probability $P_L$ that a request goes to janitor is

*Theorem 5.* The probability $P_R$ that a route discovery succeeds in JBR protocol is

$$P_R = 1-(1-P_0)^K (1-P_0)^{KE_N}$$

*Proof.* If k hops are counted in the case of a failure of route discovery without asking the janitors, then the probability that self diagnosis fails is,

$$P_{F_0} = (1-P_0)^K$$

Again, the probability that total number of $E_N$ janitors also fails to discover the route is,

$$P_{F_1} = (1-P_0)^{KE_N}$$

Then, the probability that the route recovery succeeds is,

$$P_R = 1-P_{F_0}P_{F_1} = 1-(1-P_0)^K (1-P_0)^{KE_N}$$

*Theorem 6.* The probability that a packet is successfully routed by our protocol is,

$$P_S = (1-P_{F0}^{1/k})^{E_L} + (E_L-k)(1-P_{F1}^{1/kE_N})^{(E_L-\hat{k})}$$
$$+(E_L-\hat{k})(1-P_{F1}^{1/kE_N})^{(E_L-\hat{k})}$$
$$+(1-P_B)^k[1-(1-P_0)^k(1-P_0)^{KE_N}]^{(E_L-\hat{k})}$$

*Proof.* A packet arrives in one attempt if it passes along all links without being resent by the original host again. That means that an error does not occur along the whole route and an error occurs in one link and the recovery mechanisms are launched and give the result. Therefore, we have,

$$P_S = (1-P_{F0}^{1/k})^{E_L} + (E_L-k)(1-P_{F1}^{1/kE_N})^{(E_L-\hat{k})}$$
$$+(E_L-\hat{k})(1-P_{F1}^{1/kE_N})^{(E_L-\hat{k})}$$
$$+(1-P_B)^k[1-(1-P_0)^k(1-P_0)^{KE_N}]^{(E_L-\hat{k})}$$

## 5. Performance simulations

To understand and evaluate the effectiveness of our Janitor Based Routing (JBR), we have used a detailed simulation model of the OMNeT++ simulator [8] for application layer, network layer protocols, MAC layer and physical layer models. Our mobility uses a random waypoint model by choosing a value between 0 to the maximum number of nodes present currently as its destination. The dimensions of the fields are 1300m× 1300m with 50 nodes and 100 nodes.

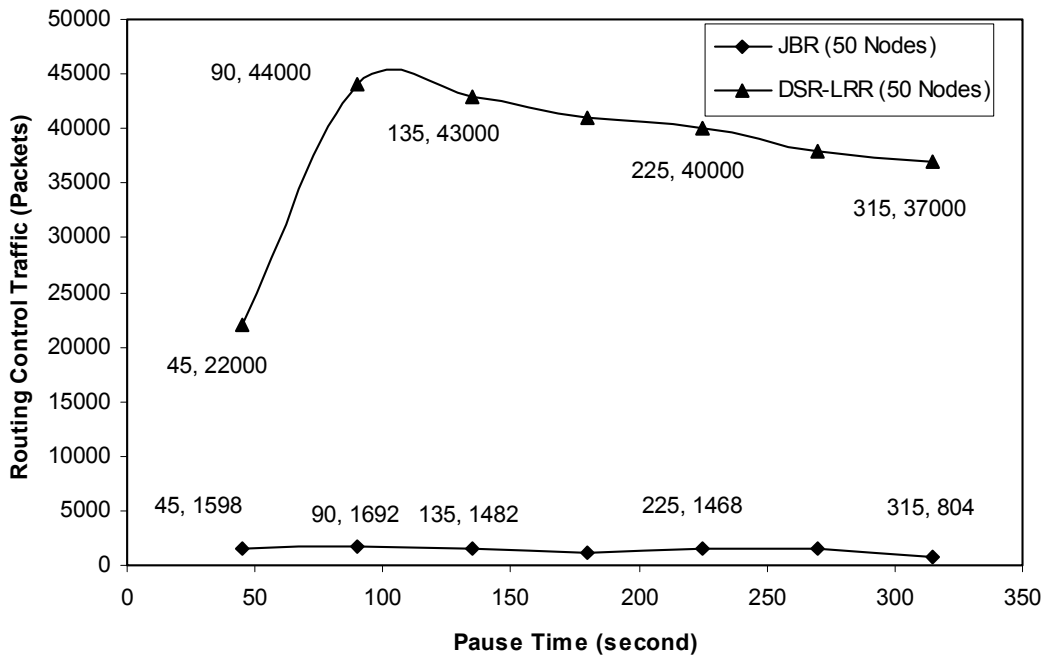

**Figure 7: Packet overhead versus pause time**

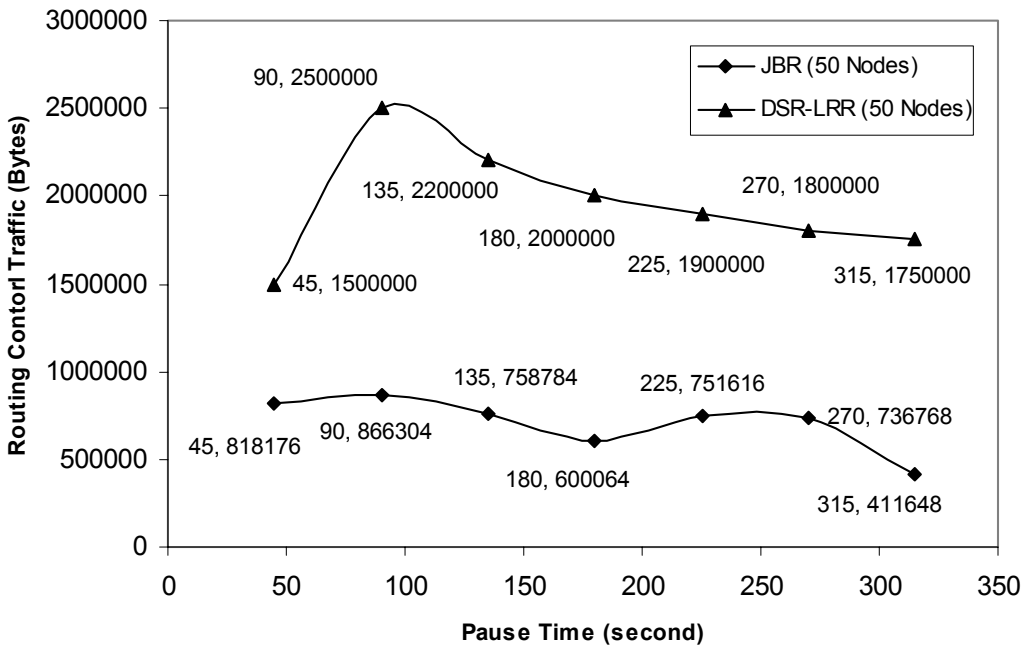

**Figure 8: Byte overhead versus pause time**

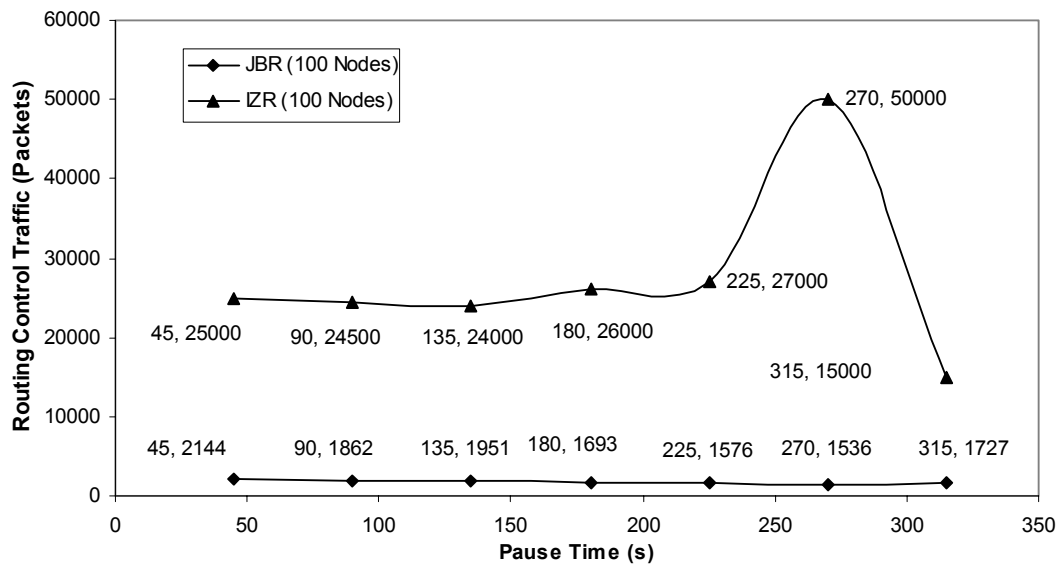

**Figure 9: Packet overhead versus pause time**

We compare the packet overhead (Fig. 7) as well as byte overhead (Fig. 8) against LRR [7] protocol which has DSR as underlying protocol. We find that our approach performs better. In case of packet and byte overhead, the highest rise in JBR routing overhead was due to the "janitor alive request" packet which is even lower in number than average number of packets in LRR-DSR. We also compare the packet overhead with IZR [9] packet overhead and observe that our protocol achieves a substantial amount of savings in packets (Fig. 9). In this case also, the highest rise of packet in JBR is lower than average number of packets in IZR. So, in comparison with LRR and IZR, JBR drastically cuts down the number of packets hence the amount of flooding experienced a steep fall. From the simulation performance, it is vivid that the performance enhancement of the ad hoc networks was clearly achieved by reducing the number of control packets formidably.

## 6. Concluding remarks

In all reactive protocols, path determination and establishment depend upon extensive flooding that influences their effectiveness. We present a Janitor Based Routing (JBR) protocol which provides a route on the fly as soon as it is broken and eliminates the need for network wide flooding. Our JBR protocol resulted in an enormous reduction in packet and byte overhead when simulated on an event driven packet level simulator. Our protocol improves other network characteristics as well. JBR is scalable and adaptable routing protocol. Mutual interaction between the constituent protocols of the JBR framework may provide useful information which can be exploited for better performance in the near future.